\documentclass[twocolumn,showpacs,aps,prl]{revtex4}

\usepackage{graphicx}%
\usepackage{dcolumn}
\usepackage{amsmath}
\usepackage{latexsym}

\begin {document}

\title
{
Self-organized critical earthquake model with moving boundary
}
\author
{
S. S. Manna and K. Bhattacharya
}
\affiliation
{Satyendra Nath Bose National Centre for Basic Sciences
    Block-JD, Sector-III, Salt Lake, Kolkata-700098, India
}
\begin{abstract}

   A globally driven self-organized critical model of earthquakes with conservative dynamics 
   has been studied. An open but moving boundary condition has been used so that the origin (epicenter)
   of every avalanche (earthquake) is at the center of the boundary. As a result, all 
   avalanches grow in equivalent conditions and the avalanche size distribution obeys finite 
   size scaling excellent. Though the recurrence time distribution of the time series of 
   avalanche sizes obeys well both the scaling forms recently observed in analysis of the 
   real data of earthquakes, it is found that the scaling function decays only exponentially 
   in contrast to a generalized gamma distribution observed in the real data analysis.
   The non-conservative version of the model shows periodicity even with open boundary.

\end{abstract}

\maketitle

      Because of the devastating effects of earthquakes on human life and wealth, understanding 
   the properties, behavior and statistics of earthquakes as well as their predictions continue 
   to remain a challenge to scientists. Over a long time attempts have been made to explain the 
   earthquake dynamics as a scale invariant process. For example, Gutenberg-Richter distribution 
   law for the earthquake magnitudes \cite{Guten}, Omori's law for the frequencies of after shocks 
   \cite {Omori} as well as recent analysis of recurrence time distributions \cite {BCDS,BCDS1,
   Corral,Corral1}, fractal distribution of epicenters \cite {Kagan,Okubo}, power law distribution 
   of the spatial distances between epicenters of successive earthquakes \cite {Paczuski}, and
   associating a scale-free network with the temporal behaviour of earthquakes \cite {Paczuski1}, all support 
   the view point that earthquakes are indeed scale invariant. On the other hand theoretically, 
   the well known Burridge-Knopoff (BK) model views the slow creeping of the continental plates 
   along the fault lines as a stick-slip process \cite {BK}. About two decades ago, Bak et. al. 
   while introducing the idea of Self-Organized Criticality (SOC) suggested that the phenomenon 
   of earthquakes may be looked upon as a SOC process since there is nobody to control the nature 
   to generate long range spatio-temporal correlations or scale-invariance \cite {BTW,Bak-Tang}.

      In this paper we study a SOC model of earthquakes and present numerical evidence to argue
   that within the frame-work of this model the earthquake dynamics is indeed scale-invariant. 
   In particular, we show that the two recently used scaling procedures for analyzing the real 
   data of earthquakes work well for our model. 

      The well known Gutenberg-Richter (GR) law says that the
   number of earthquakes $N(m)$ of magnitude at least $m$ decays exponentially with $m$ as:
\begin {equation}
\log N(m) = c_1 - c_2m.
\end {equation}
   On the other hand magnitude of an earthquake varies logarithmically with the amount of energy released:
   $\log E(m) = c_3 + c_4m$.
   Eliminating $m$ one gets, $\log N = c_1 - (c_2/c_4) \log E + (c_2c_3)/c_4$. This implies that the
   cumulative number $N(E)$ of earthquakes of energy at least $E$ decays like a power law as:
\begin {equation}
N(E) \propto E^{-b}
\end {equation}
   where $b = c_2/c_4$. Therefore the probability density of earthquakes varies as:
   ${\rm Prob}(E) \propto dN(E)/dE \propto E^{-1-b}$. Another important empirical observation is the
   Omori law which states that the frequency of after shocks decays with time as a power law:
   $D(t) \propto t^{-\gamma}$.

   Let the earthquakes be measured
   with an accuracy $m_c$ so that all earthquakes of magnitude greater than $m_c=\log_{10}(s_c)$ are detected 
   and let them be ordered in a time sequence so that the $i$-th earthquake occur at the time 
   $t_i$. The recurrence time is then defined as $\tau_i = t_i - t_{i-1}$. Bak et. al. 
   analyzed the real data of the earthquakes occured in Southern California by dividing this 
   region into a grid of cell size $L$ degrees. Considering main events, after shocks and fore shocks 
   on the same footing Bak, Christensen, Danon and Scanlon (BCDS) claimed that the recurrence time 
   follows an universal scaling function \cite {BCDS}
\begin {equation}
   {\rm Prob}(\tau,L,s_c) \sim \tau^{-\gamma}{\cal F}(\tau \frac {L^{d_f}}{s_c^{b}})
\end {equation}
   where $b$, $\gamma$ and $d_f$ are the GR exponent, Omori exponent and the fractal dimension
   of the distribution of epicenters and ${\cal F}(x)$ is an universal scaling function.
   The scaling factor $s_c^{b}/L^{d_f}$ is the mean recurrence time for the earthquakes
   having sizes at least $s_c$ which originated from a cell of size $L$.
   On the other hand Corral used a single parameter $R$ for scaling, which is the
   rate of occurrence of the earthquakes
   \cite {Corral,Corral1}:
\begin {equation}
   {\rm Prob}(\tau,R) \sim R {\cal G}(R {\tau})
\end {equation}
   where ${\cal G}(x)$ is another universal scaling function having the form of
   a generalized Gamma function. 

\begin{figure}[top]
\begin{center}
\includegraphics[width=6.0cm]{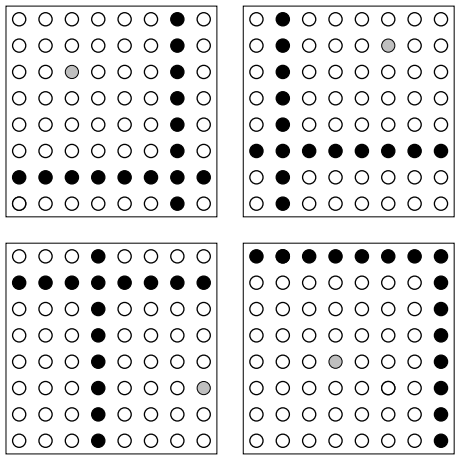}
\end{center}
\caption{
Four examples show the positions of the avalanche origins (shaded circle) and 
the corresponding boundaries sites (filled circles) on a $8 \times 8$ square lattice.
}
\end{figure}

   Bak and Tang
   devised a SOC model of earthquakes by studying a simpler version of the two-dimensional 
   BK model \cite {Bak-Tang}. The essential simplification is
   in treating the accumulated local force as a scalar as well as considering the two-dimensional
   system of blocks located at fixed positions at the sites of a regular lattice like a discrete
   space-time but continuous spin cellular automaton. 

      Olami, Feder and Christensen (OFC) studied the non-conservative version of the SOC
   model of earthquakes \cite {OFC}. Every site of a square lattice is assigned a continuous variable $f$ 
   representing the accumulated local force at that site. The system is globally driven, 
   implying that in the inactive state of no avalanches (earthquakes) the forces at all sites
   increase steadily with an uniform rate. A threshold value $f_c$ of the forces exists for 
   the stability of all sites. A site relaxes with probability one when $f_{i,j} \ge f_c$. 
   In a relaxation the force at the site is reset to zero and $\alpha$ fraction of the force
   is transmitted to each neighbor:
\begin {eqnarray}
{\rm If,} & \hspace*{0.3 cm} f_{i,j} \ge f_c, \hspace*{0.2 cm} {\rm then} \hspace*{0.2 cm} f_{i,j} \to 0
            \hspace*{0.4 cm} {\rm and} \nonumber \\
          & f_{i\pm1,j\pm1} \to f_{i\pm1,j\pm1} + \alpha f_{i,j}.
\end {eqnarray}
   Consequently,
   forces at some of the neighbors may exceed the threshold and in turn they also relax - thus a cascade
   of site relaxations propagates in the system, causing an avalanche.
   The parameter $\alpha$ varies continuously within the range $0 < \alpha \le 1/4$ \cite {OFC}. 
   The dynamics in OFC model
   is conservative for $\alpha=1/4$ and non-conservative of $\alpha < 1/4$.
   A critical value of $\alpha_c$ such that the system is in a sub-critical state for $\alpha < \alpha_c$ 
   and in a critical state for $\alpha > \alpha_c$ has been suggested for $\alpha_c \approx 0.05$ \cite {OFC},
   around 0.20 \cite {Grassberger}, $= 1/4$ \cite {Prado} and a multifractal scaling in
   \cite {Lise}.

\begin{figure}[top]
\begin{center}
\includegraphics[width=6.5cm]{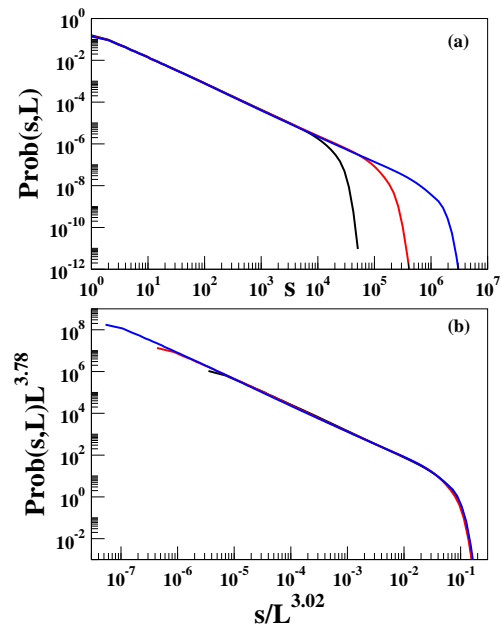}
\end{center}
\caption{(Color online)
The avalanche size distributions for three different system sizes $L = 64$ (black), 128 (red) and 256 (blue)
have been plotted on a double logarithmic scale in (a). The finite size scaling of the same data is
shown in (b).
}
\end{figure}

      We argue that assigning a fixed boundary in the SOC models of earthquakes 
   is rather artificial. In nature there is no fixed boundary for the earthquakes which absorbs 
   earth's vibrations, the seismic waves propagate in all directions till they slowly damp out at
   long distances. Presence of a fixed boundary introduces a strong non-uniformity in the system
   i.e., all measurable quantities show strong dependence on the distance from the boundary. This effect is 
   present in both conservative as well as non-conservative versions of the OFC model, but it is so strong in the
   latter case that even arriving at the stationary state becomes very difficult \cite {Grassberger}.
   It is therefore desirable that all avalanches are on the same footing
   with respect to the boundary and at the same time the origin of the avalanche should be at the farthest interior 
   point of the system.
 
      This argument prompted us to formulate a new boundary condition. Here, a globally fixed set of lattice sites
   does not constitute the boundary for all avalanches. In contrast, boundaries are different for different avalanches
   depending on the positions of the avalanche origins, and its position is constantly moved from
   one avalanche to the other. 
   
      First we make the square lattice periodic in both directions to get the topology of a torus.
   An arbitrary random distribution of forces $f_{i,j}$, drawn from a set of independent and
   identically distributed random numbers within $\{0,1\}$ are assigned at all $L^2$ sites. 
   The maximum force $f_{max}$ among all $L^2$ sites is found to be 
   at some specific location $(i_o,j_o)$ and the difference from the threshold force is
   estimated: $\delta = f_c - f_{max}$. Forces at all sites are then enhanced by
   $\delta$ so that at the origin $(i_o,j_o)$ the force reaches the threshold $f_c$.
   The avalanche then starts from the origin and a cascade of relaxations propagates away
   from the origin. 

      Now, for this avalanche, we select a specific set of lattice sites as the boundary
   such that the origin is at the center position with respect to these boundary sites.
   More precisely, on a $L \times L$ square lattice and with respect to the origin located at $(i_o,j_o)$ 
   the boundary sites form two transverse rings on the torus defined by
   one column and one row of lattice sites as (Fig. 1):
\begin {eqnarray}
   i & = & i_o+L/2 \hspace*{0.4 cm} {\rm mod}(L) \hspace*{0.4 cm} {\rm and} \nonumber \\
   j & = & j_o+L/2 \hspace*{0.4 cm} {\rm mod}(L).  
\end {eqnarray}
   When a site adjacent to the boundary relaxes, it transfers $\alpha f_{i,j}$ force to
   every non-boundary neighbor but no force to the neigbor on the boundary. Therefore corrseponding to each
   boundary neighbor $\alpha f_{i,j}$ disappears from the system and in this way the
   system looses force.

      Since the system is otherwise periodic in all directions all lattice sites are equivalent. 
   Consequently all avalanches are also equivalent since all of them grow in similar surroundings.
   In a way this is similar to elimination of surface effects in a finite size system.
   Surface profiles for the averaged force per site $\langle f \rangle$, number 
   of avalanche origins at each site $\langle e \rangle$ and average size of the avalanche per site
   $\langle s \rangle$ show uniform flat surfaces but within a very small fluctuation for all sites 
   within the lattice $L \times L$.
   We are also studying other numerically challenging problems of
   SOC using moving boundary condition.

\begin{figure}[top]
\begin{center}
\includegraphics[width=7.5cm]{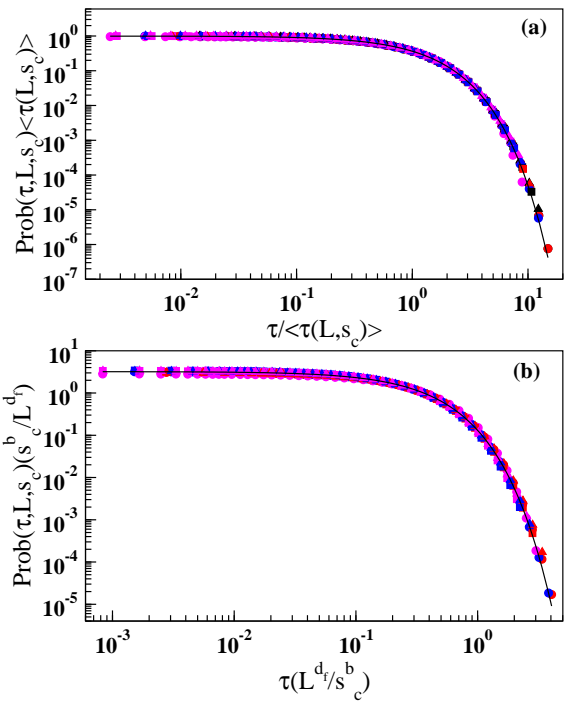}
\end{center}
\caption{(Color online)
Scaling of the waiting time distribution by the (a) Corral method (b) BCDS method.
Symbols used for: $L =$ 64 (circle), 128 (square), 256 (triangle) and for
$s_c =0$ (black), $8$ (red), $64$ (blue) and $512$ (magenta). 
Values of the scaling exponents used in (b) are $d_f$=1.67 and $b$=0.29.
The continuous line is the best fit by the functional form in Eqn. (8).
}
\end{figure}


      Since in a single relaxation, the force at the site is reduced to zero, it
   creates the possibility that more than one site (typically two) can reach the threshold 
   simultaneously.
   However, such situations occur with very low probability and in these cases we choose randomly one
   of the sites as the origin and construct boundaries with respect to this site
   but relaxation starts from both the unstable sites. Since the forces are continuously
   varying real numbers, the precision of the numbers is important as observed in \cite {Drossel}.
   To ensure that the system has indeed reached the stationary state we calculated the
   average avalanche size $\langle s(L) \rangle$ for every 10000 avalanches and monitored its variation 
   with time. This quantity first grows with time but eventually saturates. Repeating this
   calculation for different system sizes it is observed that the relaxation time grows as $L^2$. 

\begin{figure}[top]
\begin{center}
\includegraphics[width=7.5cm]{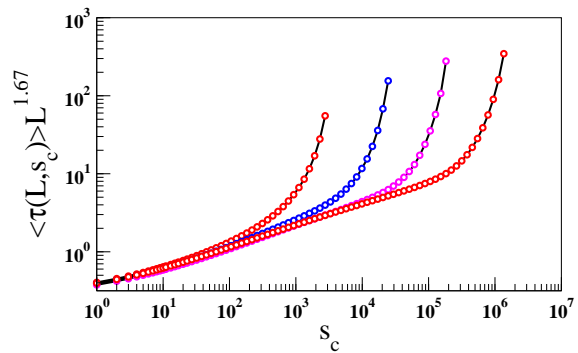}
\end{center}
\caption{(Color online)
The average recurrence time $\langle \tau(L,s_c) \rangle$ has been plotted
for different values of $s_c$ and multiplied by the system size dependent factor
$L^{d_f}$. On increasing system size the plot approaches to the variation
mentioned in Eqn. (10) with $b=0.30$.
}
\end{figure}

   First we present the results for the conservative case of $\alpha=1/4$.
   The avalanche size $s$ is the total number of relaxations in an avalanche and represents
   the total energy release in our model earthquake. ${\rm Prob}(s,L)$ is the probability that 
   a randomly selected avalanche 
   has size $s$. While for the infinitely large system size the distribution should indeed be 
   a simple power law, for the finite size systems, a finite size scaling of the distribution 
   is required:
\begin{equation}
{\rm Prob}(s,L) \sim L^{-\mu} {\cal H}(s/L^{\nu})
\end{equation}
   where the scaling function ${\cal H}(x) \sim x^{-1-b}$ for $x \to 0$ and for $x >> 1$, 
   ${\cal H}(x)$ decreases faster than a power law so that, $b=\mu/\nu-1$.
   The system size dependence of the average avalanche size and durations 
   are observed to be $\langle s(L) \rangle \sim L^{2.26}$
   and $\langle T(L) \rangle \sim L^{0.63}$. This shows that the avalanche dynamics is
   sub-diffusive. We believe that this is due to fact that force is always reset to zero in a
   relaxation which initiates more relaxations and thus increases the size
   of the avalanche.

      In Fig. 2(a) we show the plot of avalanche size distribution for three different 
   system sizes $L$ = 64, 128 and 256 on the double logarithmic scale. All of them have very 
   large portions
   of straight regions starting from very small sizes to the cut-off sizes. A scaling of the
   data with an excellent data collapse is shown in Fig. 2(b) yielding the values of
   $\nu = 3.02$ and $\mu = 3.78$ giving $b \approx 0.26$. Such a good power law behaviour as
   well as the excellent finite size scaling have been achieved only due to the moving boundary
   condition where all lattice sites as well as the avalanches are equivalent and have
   not been observed in fixed boundary cellular automata models of earthquakes before \cite {OFC,Grassberger}.

      Since we have assumed that forces at all sites increase uniformly at unit 
   rate, the time difference between successive avalanches is exactly equal to $\delta$.
   With this definition, the recurrence time distribution (RTD) ${\rm Prob}(\tau,s_c,L)$ has 
   been calculated for different system sizes $L$ as well as different $s_c$ values.
   The effects of $s_c$ and $L$ on RTD are competitive.
   For $s_c = 0$, the RTD is simply the distribution of force increments $\delta$ only.
   Since the probability of occurrence of an avalanche of size at least $s_c$ decreases
   with $s_c$, for a fixed $L$ the recurrence time increases with increasing $s_c$. On the other
   hand for a fixed $s_c$, since the maximum of the avalanche sizes increases with $L$,
   the probability of occurrence of an avalanche of size at least $s_c$
   increases with increasing $L$. Consequently the recurrence time decreases with increasing L.
   
      In Fig. 3(a) we show an unified scaling of twelve different plots with
   the minimal value of the avalanche sizes measured $s_c = 0, 8, 64$ and $512$ for three different
   system sizes $L$ =64, 128 and 256. Logarithmic binning is used for coarse-graining of the data.
   The average waiting time $\langle \tau(L,s_c) \rangle$ is calculated for each plot.
   Following Eqn. (4) we then scale every plot with corresponding $\langle \tau(L,s_c) \rangle = 1/R$
   and observe an excellent collapse of all twelve plots. This confirms the Corral scaling
   in our model. We tried to verify the Corral scaling form:
\begin {equation}
{\cal G}(x) \sim x^{-a_1}\exp(-a_2 x^{a_3})
\end {equation}
   and obtained $a_1 = 0.003$, $a_2= 1.02$ and $a_3 = 0.99$ compared to 
   $a_1 = 0.33$, $a_2= 0.63$ and $a_3 = 0.98$ observed in \cite {Corral}.
   The exponential tail in ${\cal G}(x)$ is consistent with the Gamma
   distribution observed by Corral but the observed power law decay component
   for small values of waiting times is rather absent in our model.

      To see if BCDS scaling is also valid for our model, we plotted
   ${\rm Prob}(\tau,L,s_c)(s_c^b/L^{d_f})$ vs. $\tau L^{d_f}/s_c^b $ and obtained 
   a scaling form like: 
\begin {equation}
   {\rm Prob}(\tau,L,s_c)\frac{s_c^b}{L^{d_f}}\sim {\cal F}_1(\tau \frac {L^{d_f}}{s_c^{b}}).
\end {equation}
   Here also we see a very good collapse of the nine sets of data for three system sizes
   $L=64$, 128 and 256 and for $s_c = 8, 64$ and 512. The scaling exponents
   that gave the best collapse were tuned to $d_f = 1.67$ and $b=0.29$.
   The best fit with the functional form in Eqn. (8) gives $a_1 = 0.001$, $a_2= 3.21$ and $a_3 = 0.99$
   again showing an exponential tail similar to that obtained from real data analysis \cite {Corral}
   but without any power law component.

      We therefore conclude that both the scaling forms used by Corral as well as BCDS
   are valid for scaling of the RTD data in our model.
   The scaling functions in both cases were observed to be very close to simple exponential
   decay and the power law part representing the RTD for small values of the recurrence times
   turned out to be absent. This result may also be compared with two recent analytical calculations:
   (i) a pure exponential decay of the RTD \cite {Molchan} (ii) an approximate unified law 
   compatible with the empirical observations incorporating the Omori law \cite {Saichev}. 

      For Corral's analysis it is the single parameter scaling i.e., the mean recurrence
   time $\langle \tau(L,s_c) \rangle$. However this parameter in turn also depends jointly on the
   another two competitive parameters of the distribution, namely the system size $L$
   and the avalanche size cut-off $s_c$ in the following way:
\begin {equation}
   \langle \tau(L,s_c) \rangle \propto \frac {s_c^b}{L^{d_f}}.
\end {equation}
   To check if it is really true we plotted $\langle \tau(L,s_c)\rangle L^{d_f}$
   with respect to $s_c^b$ for $L=32, 64, 128$ and
   256 using $d_f=1.67$ in Fig. 4. A nice collapse of the data for the four different system
   sizes are observed for small and intermediate values of $s_c$. Collapse of the data between two
   successive system sizes increased with the system size. The slope of the curve
   in the longest straight region corresponds to $b=0.30$. 

      Finally, we studied the OFC model using values of $\alpha < 1/4$ again on a square lattice
   of size $L$ using open but moving boundary condition. To our surprise we see that
   the dynamics become periodic after a short relaxation time of the order of $L^2$.
   This is checked by looking at the `hamming distance'. Starting from a random distribution
   of forces as before we skip some $10 L^2$ initial avalanches and store the
   force configuration in an array $f_{store}$. After that at the end of every avalanche
   we calculated the maximal site difference ${\rm max}|f_{i,j}-f_{store}(i,j)|$ and measure
   the time when this maximal difference becomes less than a small number $\epsilon = 10^{-12}$.
   The periodic time is of the order of $L^2$ but less than it, and found to depend on
   the initial distribution of force values.

      To summarize, we have studied in a model the scale invariance properties 
   observed in the real data of earthquakes over last several years by different groups.
   More specifically we studied a self-organized critical model of earthquakes
   using a square lattice cellular automaton. Using a moving boundary condition
   we could eliminate all boundary effects. We first observe that the avalanche
   size distribution of this model follow excellent finite size scaling. Further,
   the recurrence time distribution was analyzed in two ways, i.e., using Corral
   as well as BCDS scalings. We observe that our simulated data of the RTD support
   both scalings very well which leads us to conclude that the mean recurrence
   time is actually a joint function of both the system size as well as the 
   avalanche size cut-off as used to measure the waiting times. 
   
\leftline {Electronic Address: manna@boson.bose.res.in}

\end {document}